\documentclass[10pt,twocolumn,english,aps,prd,nofootinbib,eqsecnum]{revtex4}

\usepackage[utf8]{inputenc}
\usepackage{graphicx}
\usepackage{amsmath}
\usepackage{amssymb}
\usepackage{amsfonts}
\usepackage{lscape}
\usepackage{hyperref}
\usepackage{url}
\usepackage{epsfig}
\usepackage{color}
\usepackage{xcolor}
\usepackage{bm}
\usepackage{tabularx}
\usepackage{braket}
\newcommand{\be}{\begin{equation}}
\newcommand{\ee}{\end{equation}}
\newcommand{\ba}{\begin{eqnarray}}
\newcommand{\ea}{\end{eqnarray}}
\newcommand{\nn}{\nonumber}

\renewcommand{\[}{\begin{equation}}
\renewcommand{\]}{\end{equation}}

\newcommand{\OM}{\Omega_\text{M}}
\newcommand{\OK}{\Omega_\text{K}}
\newcommand{\OL}{\Omega_\Lambda}

\begin{document}

\thispagestyle{empty}

\title{Cosmic acceleration from first principles}

\author{Juan Garc\'ia-Bellido}\email[]{juan.garciabellido@uam.es}
\author{Lloren\c{c} Espinosa-Portal\'es}\email[]{llorenc.espinosa@uam.es}

\affiliation{Instituto de F\'isica Te\'orica UAM-CSIC, Universidad Auton\'oma de Madrid,
Cantoblanco, 28049 Madrid, Spain}

\date{\today}

\begin{abstract}
General relativistic entropic acceleration theory may explain the present cosmic acceleration from first principles without the need of introducing a cosmological constant. Following the covariant formulation of non-equilibrium phenomena in the context of a homogeneous and isotropic Friedmann-Lemaitre-Robertson-Walker (FLRW) metric, we find that the growth of entropy associated with the causal horizon of our universe (inside a finite bubble in eternal inflation) induces an acceleration that is essentially indistinguishable from that of $\Lambda$CDM, except for a slightly larger present rate of expansion compared to what would be expected from the CMB in $\Lambda$CDM, possibly solving the so-called $H_0$ tension. The matter content of the universe is unchanged and the coincidence problem is resolved since it is the growth of the causal horizon of matter that introduces this new relativistic entropic force. The cosmological constant is made unnecessary and the future hypersurface is Minkowsky rather than de Sitter.
\end{abstract}
\maketitle

\section{Introduction}

Most of the evolution of the universe has occurred in a state of local thermal equilibrium (i.e. without significant particle production) as the universe expands. The generally covariant formulation of non-equilibrium phenomena described in Ref.~\cite{paperI} (from now on paper I), predicts that in those periods in the evolution of the universe in which explosive particle production occurs, like in (p)reheating after inflation, or strongly first order phase transitions, or the formation of structures like galaxies and black holes, there must be a new entropic force acting on the universe, either locally or globally, depending on the phenomenon. 

In this letter we will explore the period of inevitable acceleration of the universe that ensues from the growth of entropy associated with the causal (particle) horizon. In the past, or on small scales, this acceleration is completely negligible and can be ignored. It is only in the last 6 billion years that this acceleration has started to dominate the expansion of the universe in a way that is almost indistinguishable from a vacuum energy. However, such an acceleration will eventually get diluted and the universe will end in empty flat space, not in de Sitter.

There is a neat separation between ``bulk" and ``boundary" entropic forces. The first one is due to out-of-equilibrium phenomena occurring locally, like particle production or gravitational collapse, and generates a local entropic force. The second one is due to the growth of the number of degrees of freedom in the causal boundary of space and generates a global entropic force. These forces can be added to the corresponding equations of motion, and some will be more important than others over large distances.

The driving thought behind this formulation is a simple observation. We are used to the energy conservation equation in FLRW, $D^\mu\,T_{\mu\nu}=0$, that gives rise to the continuity equation,
$\dot\rho + 3H(\rho+p) = 0\,$ where $H=\dot a/a$ is the rate of expansion of the universe. However, it can also be derived from the second law of thermodynamics, in the context of the adiabatic expansion of the universe, $T dS = d(\rho a^3) + p\,d(a^3) = 0.$ But what would happen if entropy is not conserved, in special moments of the history of the universe where non-equilibrium phenomena occurs, like the Big Bang (i.e. reheating after inflation, in the modern formulation of the theory)? In that case, $T dS\neq0$, and we find 
\be\label{eq:2LT}
\dot\rho + 3H(\rho+p) = \frac{T\dot S}{a^3}\,.
\ee
This expression, plus the Hamiltonian constraint, which links expansion, matter and spatial curvature, 
\be\label{eq:FR1}
\dot{a}^2  + k   = \frac{8\pi G}{3}\rho\,a^2\,,
\ee
gives rise to a new Friedmann/Raychaudhuri equation,
\be\label{eq:FR2}
\frac{\ddot a}{a} = -\frac{4\pi G}{3}(\rho + 3p) + \frac{4\pi G}{3}\frac{T\dot S}{a^3H}\,,
\ee
where the second term in the R.H.S. is an entropic cosmological force, which is positive for entropy production. As shown below, it is this term that will drive the accelerated expansion of the universe without the need to introduce any cosmological constant term. 

The heuristic derivation of (\ref{eq:FR2}) is obtained rigorously in the context of the generally-covariant formulation of out-of-equilibrium entropic forces in General Relativity presented in paper I, when applied to the expanding universe. In this second paper we will explore the cosmological consequences of such a formulation. In particular, we will study the cosmic acceleration induced by the causal cosmological horizon, reaching the conclusion that our present observations of supernovae dimming and a larger local rate of expansion than expected \cite{Riess:2019qba}, are a natural consequence of this general relativistic entropic acceleration (GREA) theory. 

\section{Entropic contributions to the cosmic expansion}

We described in paper I how the introduction of the entropic constraint leads to the modification of the Friedmann equations by an entropic force, and we linked this constraint to the second law of thermodynamics. 

Here we summarize our findings in paper I about the two main contributions to the thermodynamic constraint. On the one hand, one can consider the thermodynamic ``bulk'' entropy as a property of the fluid that fills the universe, whose Lagrangian is purely given by internal energy and so the matter action is written as~\cite{Mukhanov:1990me}
\begin{equation}
    S_m = \int dt \,L = - \int dt \,
    N a^3 \rho(a, S)\,,
\end{equation}
From this action one can compute also the pressure $p$, see paper I. Hence, hydrodynamic matter has also a well defined notion of temperature:
\begin{equation}
    T = - \frac{\partial L}{\partial S} = 
    - \frac{\partial \rho}{\partial s}\,,
\end{equation}
where $s$ is the entropy density. The cosmological entropic force for an expanding FLRW universe is then~\cite{paperI}
\begin{equation}
    F = - \frac{T\dot{S}}{\dot{a}} = - T \frac{dS}{da} < 0\,.
\end{equation}
Since $F < 0$, this entropic force will tend, in general, to accelerate the expansion of the universe. The various ``bulk'' entropy components could have different physical origins, like {\it e.g.} first order cosmological phase transitions or the explosive particle production during (p)reheating after inflation, which must have induced a second burst of accelerated expansion before the local fundamental interactions drive the fluid to thermodynamical equilibrium. We leave the discussion of this fascinating phenomenon for a separate publication.

On the other hand, one could also consider the effect of the ``boundary'' entropy associated to space-time itself, in particular to causal horizons. It can be incorporated in a natural way by extending the Einstein-Hilbert action with a surface term, the Gibbons-Hawking-York (GHY) term~\cite{paperI, Gibbons:1976ue, York:1972sj}. From the thermodynamic point of view, the GHY term also contributes to the internal energy of the system. Hence, it can be related to the temperature and entropy of the horizon as
\begin{equation}\label{eq:SGHY}
    S_{GHY} = \frac{1}{8\pi G}\int_{\cal H} d^3y\,\sqrt{h} K 
    = - \int dt \,N(t) \,T S\,,
\end{equation}
where we have kept the lapse function $N(t)$, to indicate that the variation of the total action with respect to it will generate a Hamiltonian constraint with an entropy term together with the ordinary matter/energy terms.

In order to illustrate this, let us now compute the GHY for two horizons of interest: the event horizon of a Schwarzschild black hole and the causal horizon of a FLRW universe. 

\subsubsection{Schwarzschild black hole}

First we consider the event horizon of a Schwarzschild black hole of mass $M$, whose space-time is described by the metric:
\begin{equation}
    ds^2 = \left(1- \frac{2GM}{r} \right) dt^2 - \left(1- \frac{2GM}{r} \right)^{-1} dr^2 -r^2 d\Omega_2^2\,.
\end{equation}
The normal vector to a 2-sphere of radius $r$ around the origin of coordinates is
\begin{equation}
    n = - \sqrt{1 - \frac{2GM}{r}} \partial_r\,.
\end{equation}
And so the GHY term (\ref{eq:SGHY}) for the event horizon, i.e. the 2-surface at $r = 2GM/c^2$, is
\begin{equation}
    S_{GHY} 
    = - \frac{1}{2} \int dt \,M c^2 = - \int dt \,T_{BH} S_{BH}\,,
\end{equation}
where $T_{BH}$ is the Hawking temperature and $S_{BH}$ \cite{Hawking:1974sw} is the Bekenstein entropy \cite{Bekenstein:1973ur} of the Schwarzschild black hole (restoring natural constants):
\begin{equation}\label{eq:TSBH}
T_{BH} = \frac{\hbar c^3}{8\pi G M}\,, \ \quad S_{BH} = \frac{4\pi G M^2}{\hbar c}\,.
\end{equation}
Let us then consider the {\em inevitable} black hole component of dark matter (independently of whether primordial black holes contribute or not with a significant fraction to cold dark matter, there is always the black hole component from stellar evolution). These black holes have a horizon embedded with a temperature and an entropy given by (\ref{eq:TSBH}). Assuming their total comoving number is conserved, their contribution to the total energy and entropy density is given by ($\hbar=c=1$)
\be
\rho_{BH} = n_{BH}\,M\,, \hspace{5mm}
s_{BH} = n_{BH}\,4\pi GM^2\,,
\ee
and therefore Eq.~(\ref{eq:2LT}) becomes
\be
a^3\frac{d}{dt}(\rho_{BH}a^3) =
T_{BH}\frac{d}{dt}(s_{BH}a^3) = 0\,,
\ee
since the number density of black holes dilutes with the volume. Therefore, the contribution of black holes of the same mass to the entropic force of the universe is negligible. Unless there are multiple black hole mergers or significant matter accretion, which change the mass or the number density of black holes, we don't expect a significant contribution to the expansion of the universe.

\subsubsection{Apparent cosmological horizon}

Let us consider here a different kind of causal horizon. A natural choice of boundary hypersurface in FLRW metric is the apparent cosmological horizon \cite{Bousso:1999xy}. We can consider a comoving sphere around the origin of coordinates $r=0$ with unit normal vector
\begin{equation}
    n = g^{rr}\partial_r = a^{-1} \sqrt{1- kr^2} \,\partial_r\,.
\end{equation}
Then the trace of its extrinsic curvature is
\begin{equation}\label{eq:Ext}
    \sqrt{h} K = 2 N(t) \,r\, a \sqrt{1 - kr^2} \sin \theta   
\end{equation}
the GHY term (\ref{eq:SGHY}) for the apparent horizon, $r_{AH}^{-1} = \sqrt{H^2-k/a^2}$, is 
\ba
    S_{GHY} &=& - \frac{1}{2G} \int dt \,N(t)\,
    H\,r_{AH}^2\nonumber \\
    &=& - \int dt \,N(t)\,T_{AH} S_{AH} \,, \label{eq:SGHY}
\ea
where $T_{AH}$ is the temperature and $S_{AH}$ the entropy associated with the apparent horizon:
\begin{equation}
    T_{AH} = 
    \frac{\hbar c\,H}{2\pi}\,, \hspace{5mm}
    S_{AH} = 
    \frac{c^3}{\hbar} \frac{\pi\,r_{AH}^2}{G}
\end{equation}
This type of ccosmological horizon does not contribute with a sufficient amount of entropy growth to affect the expansion of the universe.

\subsubsection{Causal cosmological horizon}

We turn now to the causal cosmological horizon of a FLRW universe. Let us start by considering an arbitrary comoving 2-sphere around the origin of coordinates. Then the trace of its extrinsic curvature is given by Eq.~(\ref{eq:Ext}) and the GHY term (\ref{eq:SGHY}) for the causal cosmological horizon, $d_H = a\,\eta$, with $r=\sinh(\eta\sqrt{-k})/\sqrt{-k}$, where $\eta$ is conformal time, can be written as 
\ba
    S_{GHY} &=& - \frac{1}{2G} \int dt \,N(t)\,\frac{a}{\sqrt{-k}}\sinh(2\eta\sqrt{-k}) \label{eq:GHY} \\
    &=& - \int dt \,N(t)\,T_H S_H 
    = - \int dt \,N\,a^3\rho_H\,, \nn
\ea
where $T_H$ is the temperature and $S_H$ the entropy associated with the causal cosmological horizon:
\begin{equation}
    T_H 
    = \frac{\hbar c}{2\pi}\ 
    \frac{\sinh(2\eta\sqrt{-k})}
    {a\eta^2\sqrt{-k}}\,, \hspace{5mm} 
    S_H 
    = \frac{\pi c^3}{\hbar} \frac{a^2\eta^2}{G}\,.
\end{equation}
The fact that we can naturally assign a temperature and an entropy to a hypersurface is a signal of the existence of an underlying quantum description of gravity and thermodynamics. This is made explicit by the appearance of $\hbar$ in both quantities. Their product, however, does not depend on $\hbar$ and leads to a classical emergent phenomenon, the acceleration of the universe.

\section{Accelerated expansion from entropic cosmology}

Let us consider a universe filled with ordinary radiation (high temperature photons and neutrinos), matter (baryons and dark matter) and entropy from the expanding horizon. This induces a rate of expansion that can be deduced from the Hamiltonian constraint,
\be\label{RMH}
{\cal H}^2 + k = \frac{8\pi G}{3}\,\rho\,a^2 \,,
\ee
where $\rho = \rho_R + \rho_M + \rho_H$ and ${\cal H} = a'/a= aH$ is the rate of expansion in conformal time. The first two components of $\rho$ come from the matter Lagrangian ${\cal L}_m$, and satisfy the usual dilution,
$\rho_R = \rho_R^{(0)}\,(a_0^4/a^4)\,, \ \rho_M = \rho_M^{(0)} \,(a_0^3/a^3)$, while the entropic horizon, $\rho_H = T_HS_H/a^3$, arising from the GHY surface term (\ref{eq:GHY}), satisfies the second law of thermodynamics $$\dot\rho_H + 3H(\rho_H + p_H) = \frac{T_H\dot S_H}{a^3}\,,$$ as expected for a growing entropic horizon.

\begin{figure}[h]
    \centering
    \includegraphics[width=\linewidth]{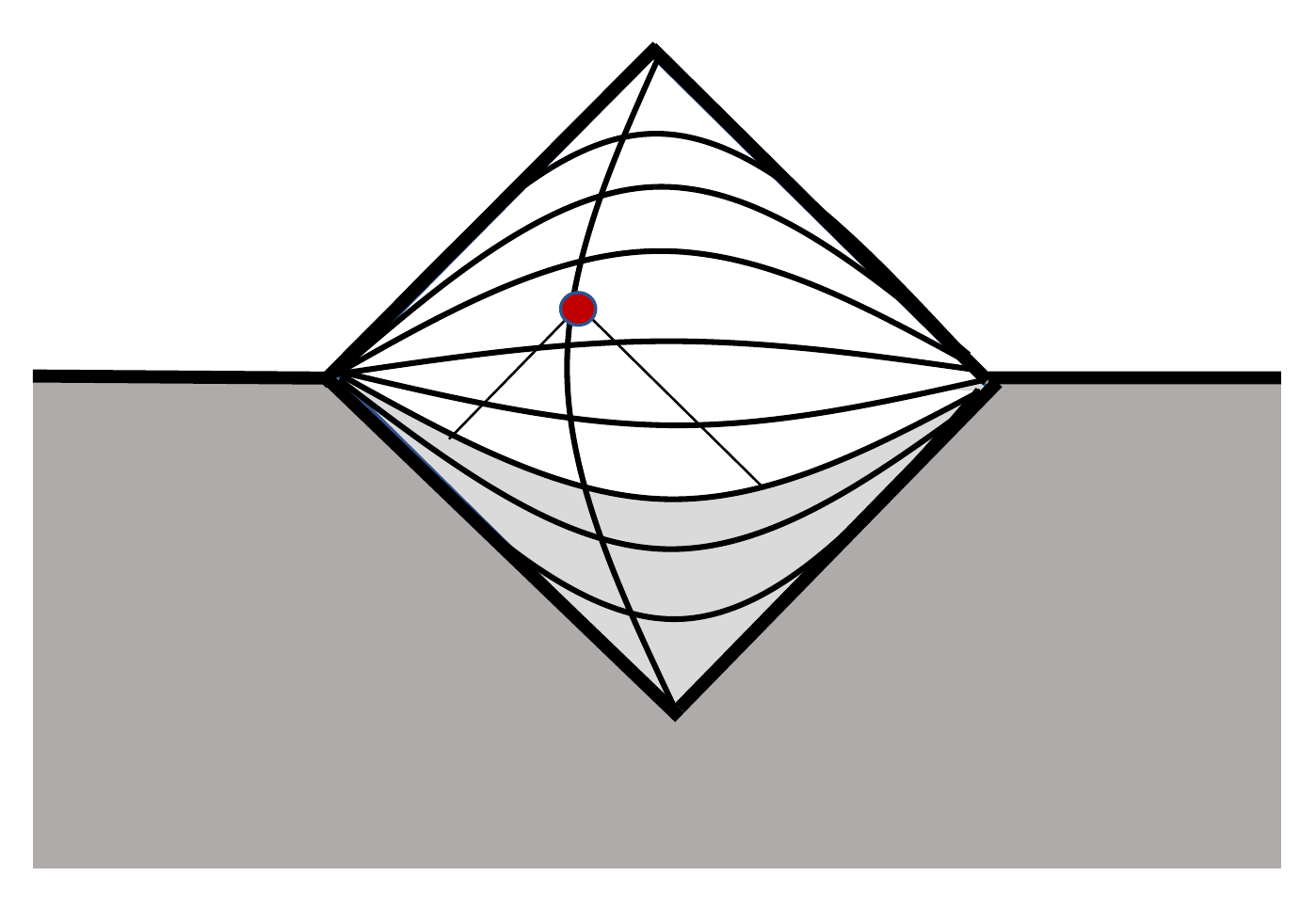}\\[5mm]
    \caption{The Penrose diagram of the scenario of open inflation described in the text. A vacuum bubble nucleated in eternal de Sitter (dark gray) has an open universe slicing inside the bubble. If enough number of e-folds occur inside the bubble, the spatial hypersurfaces are approximately flat in the past lightcone of a comoving observer (red dot). Reheating occurs at a fixed-time hypersurface (separating light-gray from white) and a FLRW universe ensues. The universe ends in Minkowsky in the far future.}
    \label{fig:Bubble}
\end{figure}

Let us then consider a causal horizon satisfying
\ba\label{eq:Causal}
\rho_H\,a^2 = \frac{T_H S_H}{a} &=& \frac{x_0}{2G}\,\sinh(2a_0H_0\eta)\,,\\[2mm]
x_0 \equiv \frac{1-\Omega_0}{\Omega_0} &=& e^{-2N}
\left(\frac{T_{\rm rh}}{T_{\rm eq}}\right)^2(1+z_{\rm eq})\,.
\ea
A possible realization of this scenario arises in the context of eternal inflation, when an open universe is nucleated in de Sitter space. If we then have enough number of e-folds of inflation ($N\geq70$) inside the true vacuum bubble, we can render the local space-time (as seen by a comoving observer) to be essentially flat ($\OK\simeq0$), although the bubble walls, and thus the true causal horizon, is at a finite coordinate distance, see Fig.~\ref{fig:Bubble}. Such an open inflation scenario is consistent with the cosmic microwave background (CMB) observations~\cite{Linde:1999wv}.

We can then solve the dynamical Friedmann equations for the accelerated universe in the context of an entropic force arising from the causal horizon (\ref{eq:Causal}). We write the Hamiltonian constraint (\ref{RMH}) in conformal time (where primes denote derivatives w.r.t. $\tau=a_0H_0\eta$) as
\ba
\left(\frac{a'}{a_0}\right)^2 &=& \OM\left(\frac{a}{a_0}\right) + 
\OK\left(\frac{a}{a_0}\right)^2  \nn \\ \label{eq:Ham}
&+&  \frac{4\pi}{3}\OK \left(\frac{a}{a_0}\right)^2 \sinh(2\tau) \,.
\ea
By solving this first order differential equation, with all cosmological parameters consistent with the CMB values (Planck 2018: $\OM\simeq0.31,\,\OK\simeq0.0006,\,h_0\simeq0.68$) and initial conditions deep in the matter era, $a_i(\tau) = a_0\,\OM\tau^2/4$, we find generic accelerating behaviour beyond the scale factor $a\sim1/2 \ (i.e.\ z\sim1)$.

\begin{figure}[h]
    \centering
    \includegraphics[width=\linewidth]{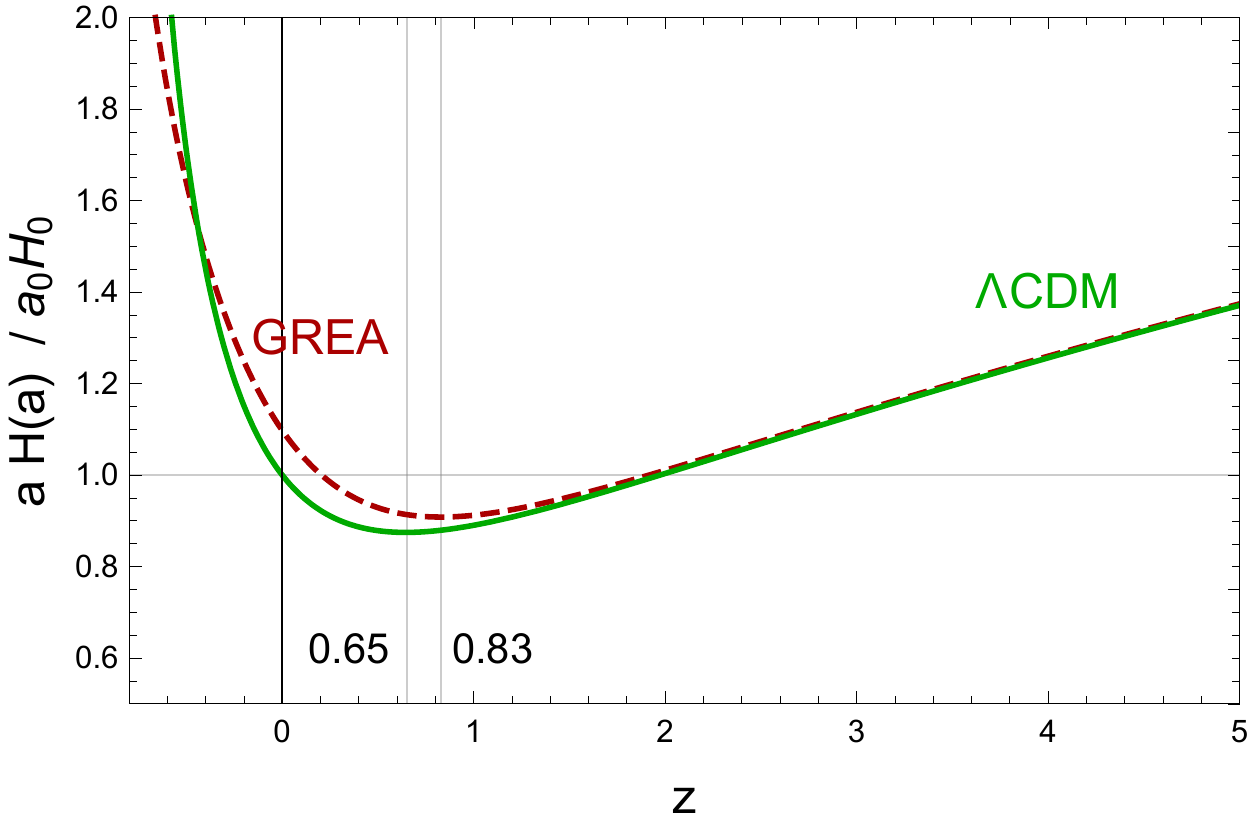}\\[5mm]
    \includegraphics[width=\linewidth]{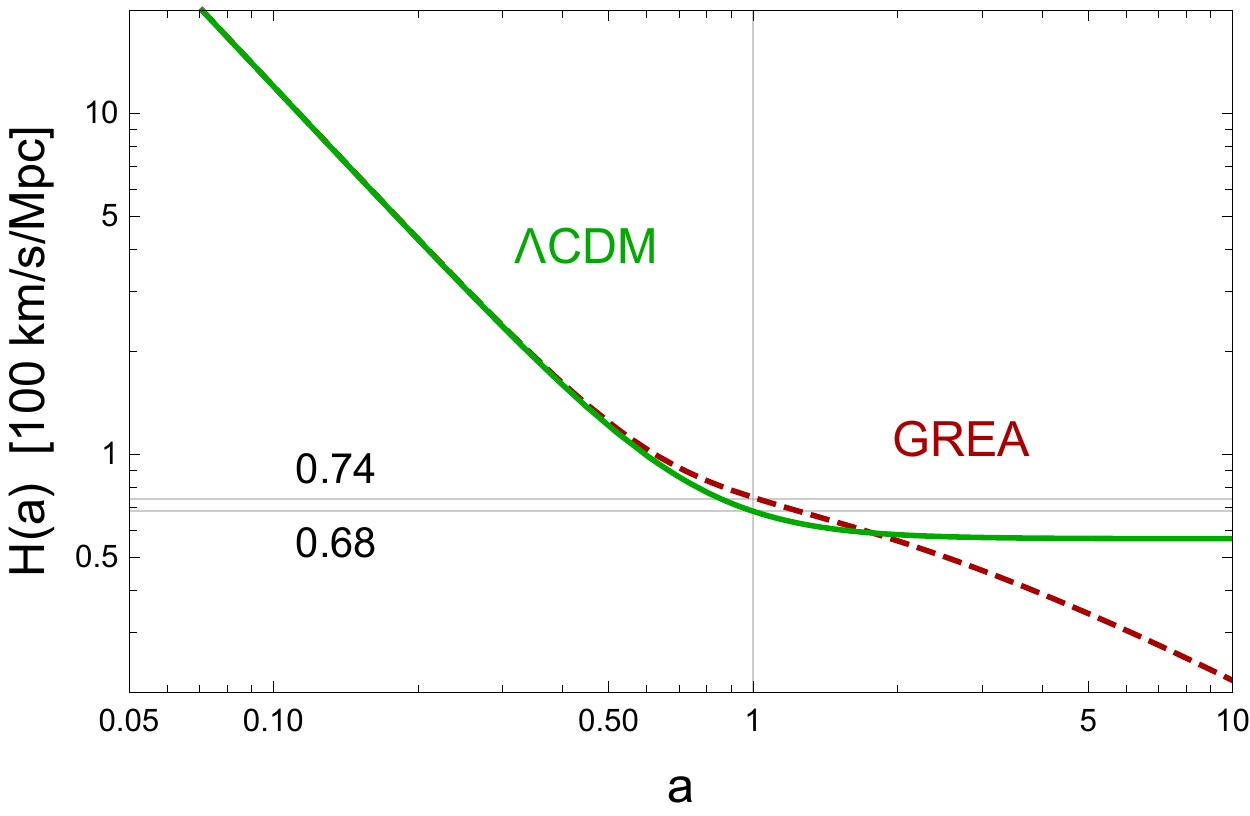}
    \caption{The upper plot shows the evolution of the inverse comoving horizon with the coasting point for each model, at $z\simeq0.65$ for $\Lambda$CDM (in green) and $z\simeq0.83$ for GREA (in red). The lower plot shows the evolution of the rate of expansion. For GREA the present rate of expansion is approximately 74~km/s/Mpc, compared with the value of 68~km/s/Mpc predicted by $\Lambda$CDM, in agreement with the asymptotic value at the CMB.}
    \label{fig:CoastingPoint}
\end{figure}

In Fig.~\ref{fig:CoastingPoint} we show the time evolution of the inverse comoving horizon ($aH/a_0H_0$) as a function of redshift $z$, for both $\Lambda$CDM and the predictions of the general relativistic entropic acceleration (GREA) theory. At large $z$, the two evolutions coincide and we recover the standard Big Bang theory at photon decoupling. However, at low redshift, $z<1$, the two differ significantly. Both evolve from matter domination, with decelerated expansion, into cosmic acceleration, going through their respective coasting points at $z=(2\OL/\OM)^{1/3}-1\simeq0.65$ and $z\simeq0.83$. The difference is still unresolved by cosmological observations~\cite{Abbott:2021bzy, Planck:2018vyg}.

Moreover, the causal horizon today is essentially identical to that in $\Lambda$CDM, {\em i.e.} $\tau_0=a_0H_0\,\eta_0\simeq3.26$, which gives $\Omega_H = \rho_H^{(0)}/\rho_c = (4\pi/3)\OK\sinh(2\tau_0) \simeq 0.71$, thus providing a reason for both the inferred value of this alternative ``dark entropy" today,
and a natural resolution of the coincidence problem: in the past such an entropic acceleration would be negligible.

Furthermore, while the rate of expansion in $\Lambda$CDM tends to $H_0=68$~km/s/Mpc, in the GREA theory the value is approximately $H_0=74$~km/s/Mpc, see Fig.~\ref{fig:CoastingPoint}, which could explain why recent local measurements have indicated a larger rate of expansion today than expected by the Standard Model of Cosmology based on $\Lambda$CDM. This could solve the so-called $H_0$ tension~\cite{Riess:2019qba}.

Note that in $\Lambda$CDM one imposes the asymptotic value of the rate of expansion at recombination and this automatically determines its present value $H(a=1) = H_0 \sqrt{\OM + \OL} = H_0$. On the other hand, in GREA we impose the same value at the CMB, but its propagation to the present time depends on $\OK$, $H(a=1) = H_0 \sqrt{\OM + \OK + 4\pi/3 \,\OK \sinh[2\tau_0]} = 1.088 \,H_0$. For a different $\OK$ parameter one obtains a different value for $H(z=0)$, thus $\OK$ is a free parameter that can be chosen to solve the Hubble tension.

\begin{figure}[h]
    \centering
    \includegraphics[width=\linewidth]{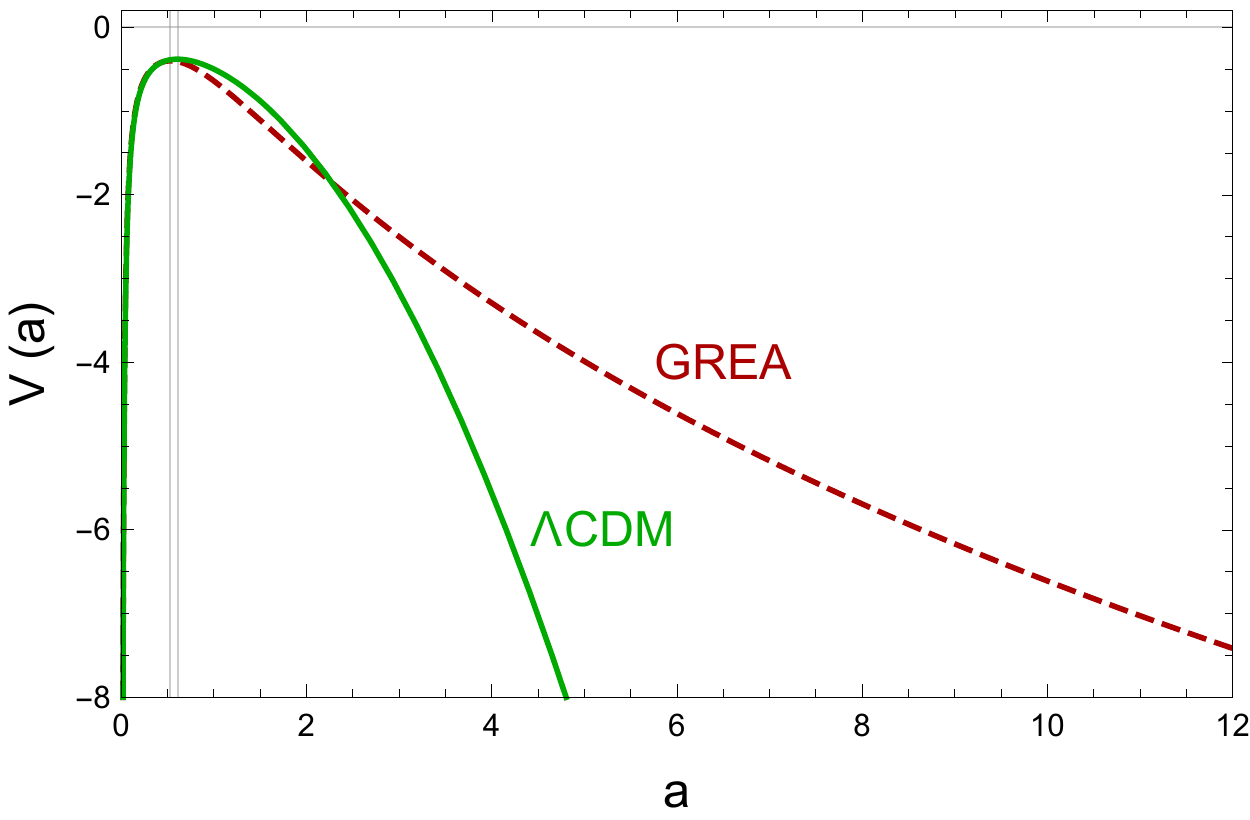}
    \caption{The effective potential for the scale factor $a$ in $\Lambda$CDM and in GREA. Note that the maximum of the potential (defining the coasting point, in vertical gray lines) occurs almost at the same time in both theories. However, the asymptotic behaviour at large scale factors is very different.}
    \label{fig:EffPotential}
\end{figure}

In Fig.~\ref{fig:EffPotential} we show the effective potential derived from the Hamiltonian constraint, $\dot a^2/2 + V(a) =$ const., both for $\Lambda$CDM, with $V(a) = - GM/a - \Lambda a^2/6$ and for GREA theory, with $V(a) = - GM/a - (4\pi G/3) \, TS(a)/a$. It is worth pointing out that the asymptotic evolution of the GREA theory is very different from that of $\Lambda$CDM. While in the latter there is a constant vacuum energy that will inevitably dominate any future dynamics, in GREA the entropic energy associated with the causal cosmological horizon will eventually be diluted, ending in empty Minkowsky space, rather than de Sitter, see Fig.~\ref{fig:Bubble}.

\section{Discussion and Conclusions}

We have applied the relativistic covariant formulation of non-equilibrium phenomena developed in paper I to the late expansion of the universe. By considering the GHY entropy associated with the causal cosmological horizon we have found a non-trivial cosmic acceleration that behaves until today very similarly to a constant vacuum energy. The relativistic entropy gradients 
generate a force that could be responsible for the observed cosmic acceleration. Rather than having a {\em constant} driving accelerated expansion, we recover the variational-principle concept of force arising from spatio-temporal gradients, in this case of entropy rather than energy. This fundamental difference provides a very simple way of testing this theory, by searching for variations in time of such a new component.

\begin{figure}[t]
    \centering
    \includegraphics[width=\linewidth]{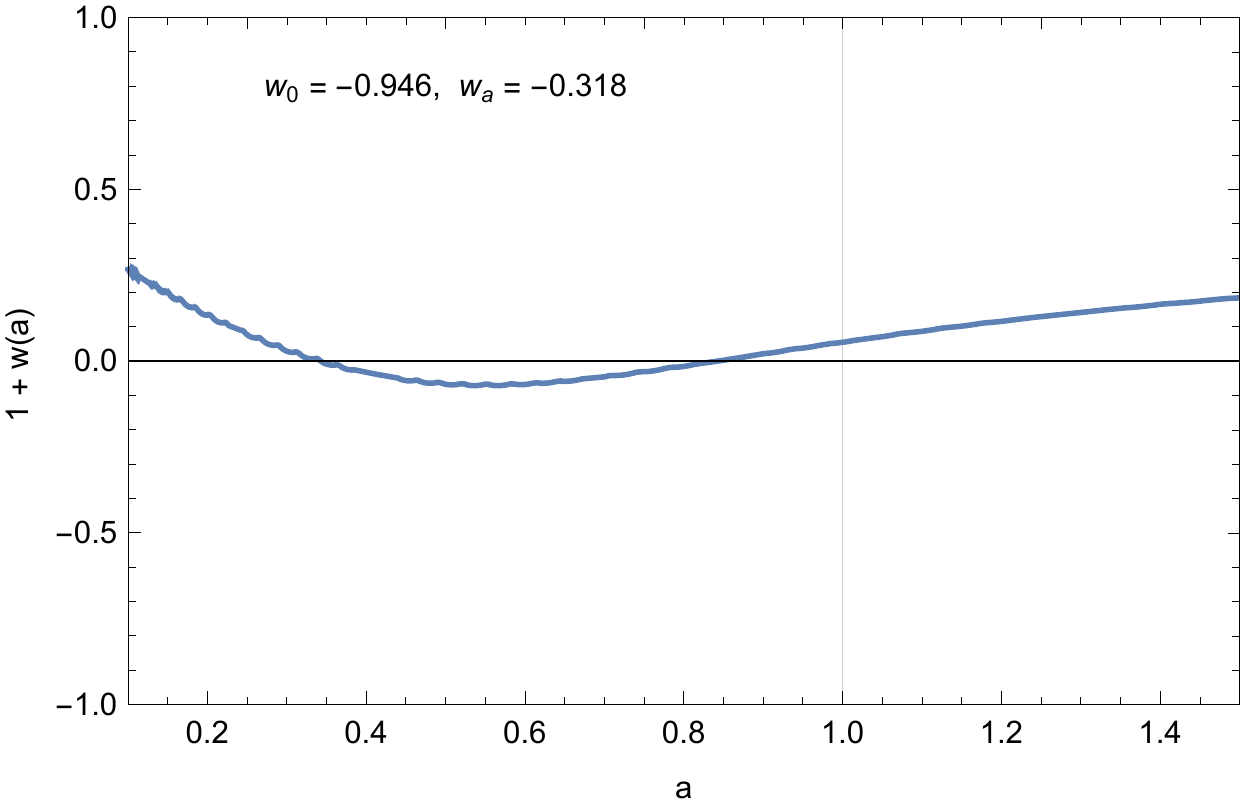}
    \caption{The effective equation of state, $w(a)=w_0+w_a(1-a)$, for this new entropic component responsible for the present acceleration, as a function of the scale factor. }
    \label{fig:wp1}
\end{figure}

We show in Fig.~\ref{fig:wp1} the effective equation of state parameter for this new entropic component, defined as
\be
1+w(a) = \frac{-1}{3}\frac{d}{d\ln a}\left[\frac{H^2(a)}{H_0^2} - 
\OM\!\left(\frac{a_0}{a}\right)^3\!- 
\OK\!\left(\frac{a_0}{a}\right)^2\right]\,.
\ee
While at present it is consistent with all observational constraints~\cite{Abbott:2021bzy, Planck:2018vyg}, in the future a precise determination of this parameter~\cite{Amendola:2016saw} may be used to test the GREA theory.~\footnote{
There are some proposals in the literature for late cosmology models trying to address the present tensions, but non of them are satisfactory from the observational point of view~\cite{Anchordoqui:2021gji}.}

Let us stress that we have derived this entropic force from first principles, by combining the Lagrangian formulation of non-equilibrium phenomena with the requirement of general covariance. In the context of a homogeneous and isotropic universe, this has provided a natural extension of the Friedmann/Raychaudhuri equation (\ref{eq:FR2}), see also paper I. When applied to the causal cosmological horizon, with Hawking-Gibbons entropy and temperature, it generated an extra entropic term to the Hamiltonian constraint and a net positive acceleration, without the need to introduce a cosmological constant nor other exotic matter. We also note that, while the temperature and entropy of the Hawking-Gibbons surface term is intrinsically quantum (there is an explicit $\hbar$ dependence), the actual entropic force is classical, it remains when we tend $\hbar\to0$, and therefore, we can argue that this basic result can be regarded as an emergent gravitational phenomenon. 

Moreover, while the inclusion of a cosmological constant, or something very similar to it, seemed to be needed by the observation of cosmic acceleration, it had always been challenging to explain the tiny value it had to have, in order to agree with observations. In our case, the small value is directly related to the size of the causal horizon at present and the local spatial curvature.\footnote{A connection between the causal horizon and the actual value of the effective cosmic acceleration today was also obtained by E.~Gazta\~naga, in a completely different scenario~\cite{Gaztanaga:2021bgb}.} 
Without the need of a cosmological constant having physical effects, one can simply put it to zero and postulate it to be protected by a more fundamental, yet unknown symmetry. We make no claim regarding the nature of this symmetry. However, the GREA theory makes this problem independent of the explanation of the accelerated expansion of the universe.

Furthermore, the coincidence problem is eliminated since it is precisely the growth of the causal horizon during the matter era that drives the relativistic entropic force responsible for the late acceleration. In the past, the entropy of the horizon was so small that the induced acceleration was negligible in practice. 
The coincidence problem for the value of $\Omega_\Lambda$ is somewhat replaced by another coincidence problem for the value of $\Omega_K$. There does not seem to be any dynamical attractor that brings it to the value required to explain the accelerated expansion of the universe. Its smallness, however, can be fully justified within the open inflation paradigm. We argue this to be a less concerning coincidence problem. 
Moreover, for an appropriate size of the causal horizon, one finds a value of the local rate of expansion compatible with local measurements while at the same time consistent with the measured value at the CMB, thus resolving the so-called $H_0$ tension~\cite{Riess:2019qba}.

Let us remark, however, that this may not be the only entropic force at play in the late universe. In particular, the formation of large structures like galaxies inevitably bring with it a local ordering of matter in the form of spiral arms, etc. This ``bulk" entropy production may be responsible for a local acceleration that could complement that of dark matter in generating the flat rotation curves of galaxies. 
Furthermore, in the case of the largest structures of the cosmic web, superclusters and supervoids, the associated entropy production from gravitational collapse should also be responsible for a local acceleration which could make large voids even deeper and emptier, thus explaining some recent observations of the ISW effect along the line of sight of supervoids~\cite{DES:2018nlb}.   
One may even consider the possibility that local entropic forces may add up and have a non-negligible effect on the background metric. Should that be the case, the dynamics of the accelerated expansion of the universe as described in this paper may be modified.

A full analysis of the perturbation theory associated with these entropic forces and their connection with large scale structure is left for a future publication.

Moreover, a direct consequence of entropic forces during the evolution of the universe, acting whenever there is significant entropy production, may have occurred at the moment when all the particles in the universe were produced. In the standard cosmological model, this occurs right at the end of inflation, when the inflaton potential energy decays into radiation, reheating the universe in the form of a gas of relativistic particles at high temperatures. Such a process is responsible for a fantastic growth in entropy from zero (at the end of inflation we only have the homogeneous zero-mode of the inflaton) to an entropy of the order of $10^{12}$ particles per Hubble patch at reheating. Today's universe encompasses $10^{77}$ of those patches and contains $10^{89}$ particles. Such a tremendous growth in just a few $e$-folds (or oscillations of the inflaton) necessarily must be accompanied by a correspondingly large entropic force, which could be responsible for a second burst of inflation. Note, however, that this second period may be very short, since the moment the particle content thermalizes through their mutual fundamental interactions, the entropy becomes maximal and those particles reach local thermal equilibrium, shutting down the entropic forces driving the acceleration in the first place. How many $e$-folds of {\em entropic inflation} happen during reheating is still unknown. We leave the analysis of this fascinating phenomenon to another publication.

Finally, just as an afterthought, we may consider the consequences that this relativistic entropic force provides for the Holographic principle \cite{tHooft:1993dmi, Susskind:1994vu}. Up to now, the correspondence between bulk gravity and thermodynamical degrees of freedom at the boundary were done in the context of static configurations. The generally covariant formulation of non-equilibrium phenomena (described in paper I) open the possibility to study the holographic correspondence in dynamical systems.

\begin{acknowledgements}

The authors acknowledge support from the Spanish Research Project PGC2018-094773-B-C32 (MINECO-FEDER) and the Centro de Excelencia Severo Ochoa Program SEV-2016-0597. The work of LEP is funded by a fellowship from ``La Caixa" Foundation (ID 100010434) with fellowship code LCF/BQ/IN18/11660041 and the European Union Horizon 2020 research and innovation programme under the Marie Sklodowska-Curie grant agreement No. 713673.
\end{acknowledgements}

\bibliographystyle{h-physrev}
\bibliography{paperEF}

\end{document}